\documentclass[acmsmall]{acmart}
\usepackage{multirow}
\usepackage[normalem]{ulem}
\usepackage{array}
\usepackage{soul}
\usepackage{hyperref}

\newcolumntype{F}[1]{%
 >{\raggedright\arraybackslash\hspace{0pt}}p{#1}}%
\newcolumntype{T}[1]{%
 >{\centering\arraybackslash\hspace{0pt}}p{#1}}%

\AtBeginDocument{%
 \providecommand\BibTeX{{%
 \normalfont B\kern-0.5em{\scshape i\kern-0.25em b}\kern-0.8em\TeX}}}

{
}

\begin{document}

\title[Mapping a Movement]{Mapping a
Movement: Exploring a Proposed Police
Training Facility in Atlanta and the Stop Cop City Movement through Online Maps}



\author{Camille Harris}
\email{charris320@gatech.edu}
\orcid{0009-0000-9227-1470}

\author{Clio Andris}
\email{clio@gatech.edu}
\orcid{0000-0002-8559-5079}

\affiliation{%
 \institution{Georgia Institute of Technology}
 \streetaddress{North Avenue}
 \city{Atlanta}
 \state{Georgia}
 \country{USA}
 \postcode{30313}
}

\renewcommand{\shortauthors}{Harris and Andris}



 \begin{abstract}

In 2021, the City of Atlanta and Atlanta Police Foundation launched plans to build a large police training facility in the South River Forest in unincorporated DeKalb County, GA. Residents of Atlanta and DeKalb County, environmental activists, police and prison abolitionists, and other activists and concerned individuals formed the movement in opposition to the facility, known as the Stop Cop City / Defend the Atlanta Forest movement. Social media and digital maps became common tools for communicating information about the facility and the movement. Here, we examine online maps about the facility and the opposition movement, originating from grassroots organizations, the City of Atlanta, news media outlets, the Atlanta Police Foundation, and individuals. We gather and examine 32 publicly available maps collected through the Google Search API, Twitter (now X), Instagram and reddit. Using a framework of critical cartography, we conduct a content analysis of these maps to identify the mapping technologies and techniques (data, cartographic elements, styles) used by different stakeholders and roles that maps and mapping technologies can play in social movements. Finally, we examine the extent to which these maps provide data to confirm or contradict concerns raised by grassroots organizations and local residents about the facility. We find that stakeholders and mapmakers use geo-spatial tools in different ways and likely have varied access to mapping technologies. We argue that documenting the use of maps to communicate information about a contentious project can help enumerate community positions and perspectives and we advocate for accessible mapmaking tools. We conclude by discussing the implications of accessibility of mapping technology and posting maps to social media, and share example map images that extend the geographic information systems (GIS) techniques seen in the retrieved maps.

{
}

 \end{abstract}





\begin{teaserfigure}
\begin{centering}
\includegraphics[width=\textwidth]{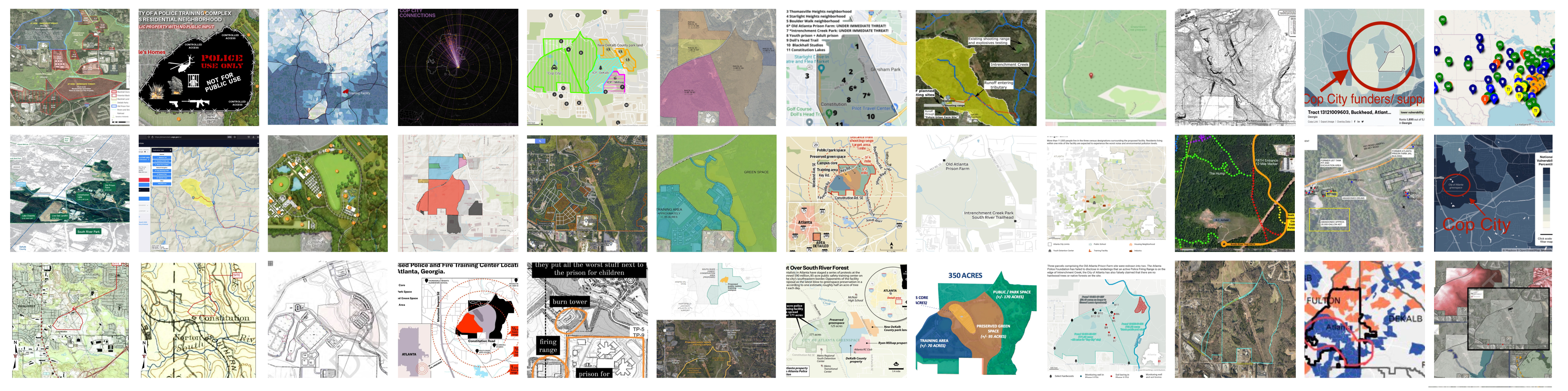}
\caption{A sample of maps retrieved from online sources that describe aspects of a proposed police training facility (colloquially known as Cop City) in Atlanta, Georgia, from proponents of the facility, the mass media, and activists who oppose the plan.}
\Description{Mosaic of maps related to the Proposed Police Training Facility}
\end{centering}
\end{teaserfigure}

\maketitle
\section{Introduction}
\subsection{Background and Motivation}

Cartographic products, including digital, print, interactive and static maps, are often used to communicate information about geographic sites where a change in the distinction, physical environment, or built environment, is planned or proposed. 
Maps play an important role in helping the public learn about the location of the proposed changes, and how changes may affect surrounding areas and people. Maps communicate information in a way that is different than graphs, because they include a representation of spatial context. Maps can communicate messages quickly and graphically, making them effective sources for mass attention \cite{eichberger2019maps, firth2014critical} that are easy to share and endorse online (e.g., via re-tweeting). 

It is well-known that the cartographic process and map products are affected by the mapmaker's purpose and can be used to execute a strategy \cite{eichberger2019maps, mackaness1995constraint}. For example, maps that explain human-environmental issues such as fracking may be viewed differently depending on the stakeholders presenting the information and the information being included or omitted. For this reason, it is valuable to study how maps are made, what they contain, and what strategies and technology are used, especially when describing a site undergoing \textit{a contentious project}. This form of analysis can capture the variation in technologies and strategies used by various stakeholders who support the project, oppose the project, or appear to remain neutral, and how these groups use technology to convey their message. 

In this work, we engage methods at the intersection of critical cartography and human computer interaction (HCI) to examine digital maps used to explain plans for a police training facility in the Atlanta, Georgia, USA, area and maps shared by members of a social movement opposing this facility. This \$90 million dollar facility, known colloquially as ``Cop City,'' is slated to be built on a 321 acre plot of forested land in the large South River Forest. Since 2021, communities of activists and residents have opposed this plan as part of the nationally-recognized, decentralized Stop Cop City and/or the Defend the Atlanta Forest Movement (which we refer to collectively as the SCC movement in this paper). The SCC Movement cites concerns such as racial justice, environmental justice, climate change, and over-policing \cite{akbar2023fight, mowatt2023people, pellow2023confronting}. Information about the project has been shared on official government channels and social media platforms, and through digital maps circulated on the web. 

\subsection{Research Questions and Contribution}
Using a research framework of \textit{critical cartography} as applied to social movements (as in \cite{eichberger2019maps, grote2021pipelines, firth2014critical}), we explore maps related to the construction of the site. Our research questions are the following:

\begin{enumerate}
 \item Which cartographic strategies, technologies, data, and map styles are used to communicate messages to the public about the training facility site? For example, do contributors producing user-generated content (UGC) use similar techniques and technologies as the mass media and government-endorsed mapmakers?
 \item To what extent do the maps include data to support or oppose claims made by activists and grassroots organizational efforts in opposition to the site (i.e., motivations of the Stop Cop City movement)?
 \item What kinds of data and messages are omitted from map representations?
\end{enumerate}

To respond to these questions, we search for maps related to the training facility and the SCC movement through search engines and social media sites. Then, through content analysis, we tag maps with commonly-used cartographic keywords sourced from cartography literature, and report on commonalities and rarities in the collection of maps. We focus our analysis on mapping and cartography, as this particular movement is site-specific. Through our findings, we provide a description of narratives, perspectives, and technological choices of community members, journalists, and government officials who create and share maps about a controversial topic.

The primary contributions of the work are as follows: First, we first present a novel corpus of cartographic images created in the context of the ongoing effort to build the controversial Atlanta Public Safety Training Facility (which we refer to in this paper as the facility) and within the movement against it. Secondly, our work leverages critical cartography and content analysis to find frequent and infrequent map elements, technologies and styles in the map corpus.We also offer discussion on use of  how map elements and data presented vary by stakeholders and the implications for this. Finally, we offer a thematic analysis of this information within the framework of motivations identified by the SCC movement in opposition to the facility, examining the extent to which concerns about the harms of the facility are directly or indirectly confirmed, disputed, or alluded to through geospatial data and map design. 


\section{Background and Related Work}
\subsection{The Proposed Police Training Facility}
Atlanta, Georgia (metropolitan area population: 4,515,419 (2020 U.S. Census)) is a city in the U.S. South that serves as the main industrial and technological economic engine for the region. Atlanta is also a historically majority Black city with significant Black history. Often called the ``Cradle of the Civil Rights Movement,'' Atlanta is home to many significant moments and locations central to the Civil Rights Movement in the 1960s, including the birthplace of civil rights figure Dr. Martin Luther King Jr., and his parish, Ebenezer Baptist Church. Racial disparities in the city remain to date, with Atlanta cited as the city with the highest income inequality and racial wealth gap in the U.S. in 2022 \cite{Jackson_2022}. 

In 2021, the City of Atlanta, together with the Atlanta Police Foundation (APF), released official plans to build a police training facility on a 321 acre plot of forested land in the South River Forest (also called the Weelaunee Forest). The site is outside, but near, the Atlanta city limits, inside DeKalb County, Georgia \cite{EIS_1}. The site's infrastructure is slated to include mock city buildings (such as a nightclub and a convenience store), animal stables, a shooting range, and administrative buildings. The project has an estimated cost of \$90 million \cite{pellow2023confronting}, and is in excess of \$248 million allocated to the Atlanta Police Department in 2024. A 2023 poll organized by polling firm \textit{Data for Progress} showed that 44 percent of Atlanta voters would likely vote for an initiative that would not allow the City of Atlanta to build the facility while 39 percent would likely vote \textit{against} such an initiative \cite{Jilani_2023, Dandekar_2023}.

The proposal was developed amid mass anti-police and abolitionist uprisings in Atlanta and across the country in 2020. This was driven nationally by the Minneapolis Police murder of George Floyd in May 2020 and driven further locally by the Atlanta Police murder of Rayshard Brooks in June 2020.

The site of the proposed facility has raised concerns for several reasons. 
The site is located in the South River Forest, which is part of Atlanta's four key sources of tree canopy, known locally as the ``lungs of Atlanta''. These forested areas are crucial for mitigating harmful impacts of climate change in Atlanta such as extreme heat, flooding, and air pollution \cite{muse2022heat, pallathadka2022urban}. Atlanta's ample tree canopy is 
also endangered \cite{evans2023perpetual} and increasing development in the area has resulted in a reduction in the canopy. 
Furthermore, the site is known for its history as a prison farm, where incarcerated individuals, predominately Black men, were forced to work from the 1920s to as recently as the 1990s \cite{CulturalStudy_1, mowatt2023people, Gant_2022}. Prior to the construction of the prison farm, the Muscogee Creek peoples were expelled from the land during the Trail of Tears in the late 1800s. The project raises concerns for continuing this legacy of exploitation of marginalized peoples, particularly of Black and indigenous populations which have been historically exploited, expelled from the area, and are at a higher risk of violence and murder by police \cite{allard2010police, byrd2017challenging}. These concerns are exacerbated by timing of the proposal for the site. Because the proposal came amid the anti-police and abolitionist uprisings, the SCC movement purports that the project will likely lead to increased police suppression of social movements
, and an increase in militaristic and deadly practices in everyday policing of marginalized Black communities. 


The movement has garnered national attention due to key escalations that took place in 2023; In January 2023, Georgia Bureau of Investigation officers killed activist Manuel Paez Teran (also known as Tortuguita) in the South River Forest \cite{mowatt2023people}. Subsequently, activists responded with street protests, vigils, and concerts. These events, collectively, resulted in the arrests of 61 individuals, who were subsequently charged with RICO\footnote{RICO stands for the Racketeer Influenced and Corrupt Organizations Act.} violations and domestic terrorism charges. In summer 2023, activists launched the ``Let Atlanta Decide'' effort, a campaign to force a legal referendum to allow Atlantans to vote to repeal the 2021 ordinance that authorized the city to lease the project site's land. ``Let Atlanta Decide'' organizers claim that the effort garnered over 100,000 signatures from Atlanta City Residents. 


Through our review of academic literature \cite{akbar2023fight}, news media coverage \cite{Brown_2023, Buchheit_2023, Pratt_2023}, statements from grassroots organizations 
in opposition to the facility, and video of public comment at Atlanta City Council Meetings \cite{mayatlcitycouncil, juneatlcitycouncil}, we identified key concerns about the facility. These include issues on the topics of environmental justice and environmental preservation, racial and economic justice, rights and well-being of local residents, police violence and over-policing, indigenous land protection, expenditure of funds collected from taxes, legal issues around land ownership, and voting rights. 
We organize the narrative of our results around these key concerns and how they are reflected in maps.



\subsection{Critical Cartography}
The study of \textit{critical cartography} focuses on subjectivity in maps with an emphasis on power dynamics \cite{Thatcher2018Cartography} to highlight who has power in the construction of maps, strategies used to communicate messages through mapping, intentional inclusion and exclusion of features, and how maps are used to justify decisions \cite{crampton2001maps, harley1989deconstructing, wood1992power, eichberger2019maps, cosgrove2008cultural}. 


Maps are not objective representations. Instead, they are highly stylized by the mapmaker, technologies, and systems the mapmaker uses. Critical cartography questions the information, visual communication methods, reasoning and strategies behind the display, and the message that map readers receive \cite{harley1989deconstructing}. A critical analysis of maps can highlight the needs of those who have little power in map construction \cite{crampton2001maps}. For example, the question of who and what is being \textit{excluded} from a map is important because maps can `mask or hide' an interest \cite{wood1992power, eichberger2019maps}. Further, as \citet{fish2023mapping} find in an analysis of map use (or lack thereof) in far right news media publications, the choice to use maps or not is a subjective decision that can be influenced by political motivations. Furthermore, many works within GIS have shown that usability challenges with mapmaking software and accessibility; in turn, software impacts map production, especially for members of the public \cite{dransch2001user, RZESZEWSKI201956, czepkiewicz2016public, jankowski2019evaluating}. When given access to mapmaking technologies, individuals have expanded capabilities to create maps that fit their needs, although this capability is limited to what is permitted within the software \cite{dransch2001user}, as well as hardware constraints, time constraints, and access to high quality data.

Critical cartography and critical GIS examines how maps and geospatial data include and exclude certain voices and perspectives \cite{elwood2002gis}. This process can be recursive, meaning that exclusion can breed more exclusion and a dominating perspective can remain so over time if unchecked \cite{thatcher2016revisiting}. 

Critical cartography recognizes that maps are reductive in part because the data visualization structures and limitations of GIS vector and raster data requires that social processes be reduced to tabular, coded, geometric representations \cite{sieber2006public}. Space for annotations and explanations of complex phenomena are limited, and thus cartographers must choose what information to include. This is especially true when making static maps (which represent more than 90 percent of the maps found in this study), that do not allow for cartographic narration \cite{caquard2014can}. When making dynamic maps, or invoking cartographic scrollytelling, a richer narrative can emerge through the inclusion of paragraphs, photos, and multiple views, but choices of tone, rhetoric, and media still empower the mapmaker.

Public participation GIS can be describe as the use of GIS to broaden public involvement in policymaking and/or leveraging GIS to promote the goals of NGOs, grassroots groups, and community-based organizations. Prior literature on public participation GIS provides further insight on critical cartography in relation to power and silencing. For instance, \citet{cidell2008challenging}, conducts a participatory GIS study of residents' challenges to sound maps through analysis of public comment recordings. They find that public comment enabled residents to challenge how data on ambient sound was collected and portrayed (as contour lines) and suggested enhanced resident involvement in the cartographic process. One critique of public participation GIS levied by Sieber is the following: ``GIS distracts grassroots groups and others from proven activist strategies such as protest and retreats from questioning the general framework of policymaking and distribution of power''\cite[p. 491]{sieber2006public}. Sieber argues that despite the growing ubiquity of GIS systems, the use of GIS technologies in grassroots organizations can lead to a false illusion of control, rather than using proven strategies to disrupt the actual dynamics of control and power. Despite this, Sieber points out that analysis of conventional models of participatory GIS do not include oppositional forms, such as uses of GIS to plan protests or riots, or other uses of GIS that explicitly challenge authority. Our work builds on this criticism of public participation GIS and analyzes oppositional use of GIS, while contrasting to institutionally aligned uses of GIS technology.


 

Maps play an important role in presenting information in social movements \cite{eichberger2019maps}, especially in describing the geographic and environmental factors surrounding social movements within a specific locale. Throughout the progression of the Stop Cop City movement, police and city officials, news media sources, grassroots organizations, community members, and other stakeholders have used cartographic images to present information about the facility. Many maps in this movement are forms of \textit{counter-mapping}, wherein mapmakers share maps to make an argument against prevailing `official' maps. Such efforts to use technology to check those in institutional power has been called \textit{sousveillance} \cite{mann2013new, Thatcher2018Cartography}. Conversely, many maps in our study can be considered `official' maps put forward by institutions of power such as the City of Atlanta and the APF.

\subsection{Online Maps, HCI, and Social Movements}
Several HCI studies include social justice concepts \cite{bellini2022there, fox2016exploring}, such as critical race theory \cite{ogbonnaya2020critical, smith2020s}, environmental justice \cite{bates2018future}, feminism \cite{10.1145/1978942.1979041, 10.1145/3025453.3025766}, disability justice \cite{sum2022dreaming, spiel2020nothing, williams2021articulations}, anti-capitalism and post-growth \cite{10.1145/3624981}, anti-gentrification and housing support \cite{corbett2019engaging}, and anti-police-violence and abolition \cite{10.5898/JHRI.5.3.Asaro, Gerber2018ParticipatorySF, carrera2023unseen}. Other works within HCI detail findings on building digital tools for supporting community organizing work \cite{irannejad2020supporting} and examining the tools currently in use by organizers \cite{soden2021dilemmas, 10.1145/2470654.2466262}. However, few works explore the role of digital maps within specific movements. Our work builds on a narrow selection of prior studies at the intersection of HCI and critical cartography that explicitly examine the role of online maps and digital mapping in social movements and community organizational efforts \cite{carrera2023unseen, eichberger2019maps, lundine2012youth}.


\textcolor{black}{Although similar work is limited in the field of CSCW,} \citet{carrera2023unseen} also engages the role of maps and mapping in community organizations and social movements by examining digital mapping as a tool in abolitionist organizing. They interview abolitionist organizers in response to the 2020 uprisings across North America and conduct content analysis of websites with information on prison and police abolition \cite{carrera2023unseen}. A key finding of the work is the opportunities for the use of mapping technology by abolitionist organizers to geographically display quantitative information that supports abolitionist arguments to the public \cite{carrera2023unseen}. In our study, we expand upon this finding, examining maps that appear to support, oppose, or remain neutral to a social justice cause, rather than only those supporting the cause, to convey differences in maps across various stakeholders. Further, our study focuses on a specific local movement providing more depth, rather than breadth across several locales.

Most similar to our work, \citet{eichberger2019maps} offers an analysis of online maps with information about the Sacred Stone Camp established by the Standing Rock Sioux Tribe to protest construction of the Dakota Access Pipeline. The work outlines key strategic mapping choices and design of maps made by indigenous organizers, scholars, news media, and corporations supporting the pipeline. \citet{eichberger2019maps} examines the information included or excluded from maps, and contrasts Western cartography practices that have been integral to colonization, with indigenous approaches to mapping the region, and emphasizing the role positionality plays in map production, a crucial principal of critical cartography. \citet{eichberger2019maps} concludes this analysis by inviting future scholars to interrogate the positionality of maps, and the impact that inclusions and exclusion of key information in mapping can have in shaping public and online discourse. In this work, we respond to this call through a content analysis of maps related to another social movement. 

Our study builds on a limited body of research at the intersection of critical cartography, human-computer-interaction, and studies of social justice and social movements. \textcolor{black}{This work helps create a stronger bridge between CSCW, GIS, and cartography, by examining mapmaker tools, media, and rhetoric using the lens of critical cartography. Critical cartography is a natural intersection because it questions who made the map, why design choices were made, who the audience is, and how the ecosystem of creators and users is affected by data availability, training, and issues such as colonialism (e.g., imposing naming conventions that erase local and indigenous names and sites). Maps can be powerful communication tools, and their reach is broadened when they are disseminated online through social media or online publications. We see an opportunity within HCI to leverage critical cartography to analyze them. In this work, we look to do so while also analyzing the implications for activists and community members in the context of a local social issue. Concretely, we contribute a corpus of map image data and provide an analysis of the corpus through a lens of HCI and critical cartography.}

\section{Data and Methods}
\subsection{Mapping Data Acquisition}
\subsubsection{Data Collection}
We collected static images of maps and interactive maps from websites and reports that we found through Google, Google Images \footnote{Google Images: \url{http//images.google.com}}, Twitter\footnote{We refer to the social media site X as Twitter in this manuscript, and the posts made on the site as tweets.}, Instagram, and reddit. We chose to analyze maps from social media to illustrate how maps may play a role in online discussions about the facility. We included results from Google Images to include maps that are likely to be viewed by users who search for the facility online. This strategy allowed us to retrieve potentially influential and well-circulated maps. To search Google, we used the Google search API with the phrases: `Cop City', `Stop Cop City', `Atlanta Public Safety Training Center', `Atlanta Police Training Facility', each followed by the word `map'. We also followed hyperlinks embedded in websites discovered through the Google Searches. For searches on Twitter and Instagram, we used the same set of search terms and used the search terms as hashtags (e.g., \#stopcopcity, \#atlantapublicsafetytrainingfacility). We searched Twitter by entering these terms into the Twitter search bar and traversing images in the Media tab and the Top tab. The Top tab returned the most popular results containing the keywords (i.e., most liked and re-tweeted), and the Media tab returned all results containing an image or video and the search term. Both searches returned results in order of recency. Finally we sourced maps from reddit by collecting images from the `r/Weelaunee' subreddit, searching for maps within the `visual art, maps, etc.' tag. These searches were conducted multiple times between May 2023 and January 2024.

We also retrieved images from authoritative documents, including an Environmental Site Assessment, Limited Site Investigation, and Cultural Resource Assessment, which we collectively refer to as environmental impact studies (EIS). Each can be found on the official APF website.\footnote{Atlanta Police Foundation, Public Safety Training Center: \url{https://atlantapolicefoundation.org/programs/public-safety-training-center}} Information from the APF website also led us to a set of documents from Community Stakeholder Advisory Committee Meetings.\footnote{City of Atlanta, Community Stakeholder Advisory Committee: \url{https://www.atltrainingcenter.com/community/community-stakeholder-advisory-committee}} The Atlanta City Council established this committee in 2021 to increase community input about the facility. We discovered maps shared at their monthly meetings by downloading slide deck presentations from their meeting website. We  include them in our corpus but do \textcolor{black}{not} include them in the analysis. 

\subsubsection{Exclusion Criteria}
\textcolor{black}{Before thematic coding all maps collected were screened by both authors to ensure all data included information about the proposed APF facility site and its geography.} Following initial data collection, we excluded \textcolor{black}{maps} that were used artistically as part of a flyer or a graphic. Unless the underlying map or the annotations themselves included unique information, we did not include these graphics. The same map image was frequently reposted by multiple sources, and we removed duplicated maps from our map corpus. Maps of the site that were created before the facility was proposed were rare, and were \textcolor{black}{also} not used in the analysis.

The data acquisition process resulted in a total of 45 unique map images, including annotated aerial photos (i.e., orthorectified imagery with graphics or freehand mouse-drawn drawings), site plans furnished by local government and contracted planners, architects, and civil engineers. We retrieved three maps from the EIS documents. We reduced this set of maps to 32 unique maps after applying the exclusion criteria described above. The full corpus of 45 maps (including a map found outside the search) and a brief description of the maps found in the EIS documents is included in the supplementary materials. Table \ref{Tab:DataSources} denotes the full set of 32 maps used for keyword analysis. \textcolor{black}{Most of the 32 maps we retrieved, including two interactive maps (rows 21 and 32 in Table \ref{Tab:DataSources}), were sourced from news articles.} 

\subsection{Map Analysis}
To analyze maps, we use a set of keywords outlined in geovisualization and cartography literature (e.g., \citet{slocum2022thematic}) such as: reference map, thematic map, choropleth map, and map elements such as scale bar, north arrow, legend and inset map, and keywords from the U.S. National Archives including topographic, aerial, relief, and land use \cite{NatArch2023}. We use keywords to denote visual variables such as: color, size, and transparency \cite{Bertin83} and reference visual features such as target shapes. We note trends in UGC methods such as annotation or screencapping. When known or easy to ascertain, we noted the type of tools, technology, data sources, and software (such as ArcGIS or Datawrapper) used to create the map, this information is in a separate column of the corpus table. \textcolor{black}{This method follows prior map content analysis methods where individuals coded elements in a large map collection using a set of keywords that describe land cover types, color, and the presence or absence of key map elements such as scale bars or contour lines \cite{muehlenhaus2010lost, gavsperivc2023new}. It also follows an analysis of crowd-sourced word counts on the topic of cartographic standards that also reported map-related keyword frequency \cite{buckley2020quantitative}.}

We apply keywords to each image using a spreadsheet and subsequently organize and clean the keywords in the R Statistical Computing Environment. We create a set of high-level themes and from those identify four top-level themes from the keywords. Then, we tally the total frequency of these keywords into tables and provide a descriptive analysis of the results and our interpretation.

\textcolor{black}{We annotated maps with 166 unique keywords and 375 total keywords (see Table \ref{Tab:LowLvlKeywords}). After keyword tagging, the unique set of keywords were organized into 26 high level themes (Table \ref{tab:ThemeKeywords1}). The most common themes were map element, color (especially red, green, and blue), area/geometry (such as polygons of the lot), roads, hydrological features, civic elements such as schools, and basemaps. The majority of keywords were map-related (e.g., `scale bar'), followed by human-related (e.g., `Blackhall Studios'), justice-related (e.g., `shooting range'), and nature-related (e.g., `trees'). Justice-related issues include the following top-level themes: Danger, Environmental Impact, Prison, and two terms that describe the facility: Cop City and Training Area. Ultimately this resulted in 261 map-related keyword appearances, 40 human-related appearances, 40 justice-related appearances and 34 nature-related appearances.}

\textcolor{black}{Each map had an average of about 11.7 keywords. We annotated maps with 166 unique keywords and 375 total keywords (Table \ref{Tab:LowLvlKeywords}) and these keywords were organized into 26 high level themes (Table \ref{tab:ThemeKeywords1})}.

\section{Results}

\small{
{
\begin{table}[]
 \caption{Data Sources for Keyword Analysis, Sorted Ascending by Publication Date}
\begin{tabular}{|p{0.05\linewidth}|p{0.32\linewidth}|p{0.35\linewidth}|p{0.11\linewidth}|} 
 \hline
\textbf{No.} & \textbf{Publication \& Author} & \textbf{Sources and Software} & \textbf{Date} \\ \hline
 1 & \href{https://atlantapolicefoundation.org/wp-content/uploads/2022/12/Phase-I-Environmental-Site-Assessment.pdf}{Terracon (page 91)} & Various sources such as USGS topographic maps and aerial imagery & Apr 22, 2021 \\ \hline

 2 & \href{https://twitter.com/defendATLforest/status/1392999833234026498}{Defend the Atlanta Forest (Twitter $@defendatlforest$)} & \textit{None stated} & May 13, 2021 \\ \hline

3,4 & \href{https://www.mainlinezine.com/research-shows-that-weapons-testing-at-new-police-training-facility-could-expose-the-public-to-toxic-chemicals-contaminate-urban-farm-and-south-river/}{Mainlinezine, By Wayne Butler} & USGS StreamStats, APF Data & Jul 14, 2021 \\ \hline
5 & \href{https://atlanta.urbanize.city/post/great-atlanta-police-fire-training-center-debate-where-do-you-stand}{Urbanize Atlanta, By Josh Green} & Atlanta Police Foundation & Aug 25, 2021 \\ \hline
6 & \href{https://www.ajc.com/news/atlanta-news/proposal-on-police-fire-training-center-passes-to-finance-committee-as-opposition-continues/XLYRAOKDR5GIXOWSYCVKUGJYJY/}{Atlanta Journal Constitution, By Anjali Huynh} & Atlanta Police Foundation & Aug 10, 2021 \\ \hline
7 & \href{https://theappeal.org/atlanta-cop-city-police-training-facility/}{The Appeal, By Aja Arnold} & \textit{None stated} & Dec 08, 2021 \\ \hline
8, 9 & \href{https://atlpresscollective.com/2022/01/23/the-atlanta-public-safety-training-center-development-team-is-misleading-stakeholders-regarding-environmental-assessments-and-avoiding-due-diligence/}{Terracon, Atlanta Community Press Collective, by Lily Ponitz} & (Reproduction of maps by) Terracon & Jan 23, 2022
\\ \hline
10-12 & \href{https://twitter.com/atlanta_press/status/1524438228169904129?s=20}{Atlanta Community Press Collective (Twitter $@atlanta$\textunderscore$press$)} & Dekalb County Property Information Website, Google Sheets & May 11, 2022 \\ \hline
13 & \href{https://atlanta.capitalbnews.org/atlanta-cop-city-climate-change/}{Capital B News, By Adam Mahoney and Adjoa Danso} & OpenStreetMap, Datawrapper & Jun 6, 2022 \\ \hline
14 & \href{https://defendtheatlantaforest.org/new-map-of-weelaunee-1/}{Defend the Atlanta Forest} & Google Maps & Jul 16, 2022
\\ \hline
15 & \href{https://www.ajc.com/neighborhoods/dekalb/forest-defenders-use-extreme-tactics-in-fight-to-stop-cop-city/CQCJHWEYWZBATPEX5ZKA6IRVYM/}{The Atlanta Journal-Constitution, By Tyler Estep, Map: Hannah Ziegler} & OpenStreetMap, Datawrapper & Aug 12, 2022 \\ \hline
16, 17 & \href{https://www.cnn.com/2022/09/24/us/atlanta-public-safety-training-center-plans-community/index.html}{CNN, By Christina Maxouris, Map: Renee Rigdon} & South River Forest Coalition, Atlanta Police Foundation, Google & Sep 24, 2022 \\ \hline

18 & \href{https://www.axios.com/local/atlanta/2023/02/01/cop-city-gets-a-green-light-atlanta}{Axios, By Emma Hurt} & City of Atlanta & Feb 1, 2023 \\ \hline
19 & \href{https://twitter.com/IsThatABananaOr/status/1622494058844237824?s=20}{Fleximillien, (Twitter $@thatabananaor$)} & WABE.org, Atlanta Regional Commission, U.S. Census Bureau & Feb 6, 2023
\\ \hline
20 & \href{https://saportareport.com/training-center-maps-suggest-outside-property-used-to-meet-green-space-requirements/sections/reports/johnruch/}{Saporta Report, By John Ruch} & Affidavit of R. Baskin, Atlanta Police Foundation & Mar 1, 2023 \\ \hline
21 & \href{https://atlpresscollective.com/2023/03/04/land-swap-or-cop-city-a-guide/}{Atlanta Community Press Collective} & OpenStreetMap, Datawrapper & Mar 4, 2023 \\ \hline
22 & \href{https://www.npr.org/2023/03/07/1161343394/atlanta-cop-city-protests-explained}{National Public Radio, By Bill Chappell} & City of Atlanta & Mar 7, 2023 \\ \hline
23 & \href{https://insideclimatenews.org/news/08032023/atlanta-cop-city-forest-justice-trees/}{Inside Climate News, By Victoria St. Martin, Map: Paul Horn} & Atlanta Community Press Collective, Atlanta Police Foundation, ESRI & Mar 8, 2023 \\ \hline
24 & \href{https://www.reddit.com/r/Weelaunee/comments/12xwfgl/the_fruits_of_liberal_reformism/}{Reddit.com/r/Weelaunee, (\textit{u/n0noTAGAinnxw4Yn3wp7})} & CAD, \textit{none stated} & Apr 24, 2023
\\ \hline

25 & \href{https://saportareport.com/dekalb-commissioner-and-petition-call-for-reopening-park-shuttered-in-protest-crackdown/columnists/johnruch/}{Saporta Report by John Rush} & South River Watershed Alliance & May 1, 2023\\ 
\hline
26 & \href{https://www.reddit.com/r/Weelaunee/comments/13mfwrl/police_training_centers_across_colonial_georgia/}{Reddit.com/r/Weelaunee, (\textit{u/n0noTAGAinnxw4Yn3wp7})} & Georgia Public Safety Training Center & May 20 2023
\\ \hline
27 & \href{https://mappingatlanta.org/2023/05/30/the-real-outside-agitators/comment-page-1/}{Mapping Atlanta, By Taylor Shelton} & QGIS & May 30, 2023 \\ \hline
28 & \href{https://twitter.com/AlexIp718/status/1686877226455752704?s=20}{The Xylom, By Alex Ip ($@AlexIp718$)} & OpenStreetMap, Datawrapper & Aug 2, 2023 \\ \hline 
29, 30 & \href{https://twitter.com/antiracistsouth/status/1708875924706316650}{Anti-Racist South, (Twitter $@antiracistsouth$)} & Environmental Defense Fund, Texas A\&M University, Darkhorse Analytics, MapBox, OpenStreetMap & Oct 2, 2023
\\ \hline
31 & 
\href{https://atlantapolicefoundation.org/programs/public-safety-training-center/}{Atlanta Police Foundation} & LS3P Architects & N.D. \\ \hline
32 & \href{https://StopCopCitySolidarity.org}{StopCopCitySolidarity.org via RAM INC (Twitter $@resistabolish$)} & Leaflet, uMap, Django, OpenStreetMap & N.D. \\ \hline
\end{tabular}
 \label{Tab:DataSources}
\end{table}
}
}

\textcolor{black}{We find that most of the} maps we retrieved were largely reference maps, i.e., maps that show the absolute and relative locations of geographic features (e.g., water, roads, administrative areas).\footnote{In comparison, thematic maps show a theme such as population density, average age per census unit, or land cover type.} Many maps had a scale bar (31.2\%), inset map(s) (28.1\%), legend (25\%) and a north arrow (18.8\%). These elements suggest that mapmakers have some background or training in cartography; this is unsurprising since much of our content is from news articles. The most common features were polygons that represented an enclosed area (37.5\%) followed by roads (34.4\%). Very few maps did not include context or a basemap (two were tagged as `floating polygons'), and many included aerial imagery (satellite photos, etc.) (25\%). Pictorial icons were rare, but seemed to be a source of creative expression, such as the clapboard on an interactive map from the \textit{Atlanta Press Collective} (Figure \ref{articulatingtheplan}C). 
\begin{table}[h]
 \caption{Top Individual Keywords with Three or More Appearances in the Coded Map Data}
\begin{tabular}{|llll|}
 \hline
\textbf{Keyword} & \textbf{Appearances} & \textbf{Keyword (cont.)} & \textbf{Appearances} \\
 \hline
Red & 12 & Footprints & 4 \\
Polygons & 12 & Variety of Uses Inside Polygon & 4 \\
Roads & 11 & Basemap & 4 \\
Scale Bar & 10 & Black & 4 \\
Inset Map & 9 & Target & 4 \\
Blue & 8 & Webtool & 4 \\
Green & 8 & Counties & 4 \\
Aerial Imagery & 8 & Government Tool & 4 \\
Labels & 8 & Trees & 4 \\
Legend & 8 & Neighborhoods & 3 \\
Screencap & 7 & Demographic Data & 3 \\
North Arrow & 6 & Youth Detention Center & 3 \\
Faded & 6 & Orange & 3 \\
Partial Transparency & 6 & Annotation & 3 \\
Parcels & 5 & CAD & 3 \\
Shaded Relief & 5 & Tracts & 3 \\
Icons & 5 & Contour Lines & 3 \\
Shooting Range & 5 & Number Labels & 3 \\
School & 4 & Choropleth Map & 3 \\
Old Prison Farm & 4 & City Boundaries & 3 \\
 \hline
\end{tabular}
\label{Tab:LowLvlKeywords}
\end{table}

{ 
\begin{table} [b]
\begin{centering}
 \caption{Count of Keywords by Coded High-Level Theme. \label{tab:ThemeKeywords1}}
 \begin{tabular}{| p{0.20\linewidth} | p{0.10\linewidth} | p{0.10\linewidth} | p{0.22\linewidth} |} 


 \hline
\textbf{Theme} & \textbf{Count of Appearances} & \textbf{Distinct Keyword} & \textbf{Example Keywords}\\
 \hline
Map Element & 59 & 13 & Area Measurement Labels \\
Color & 45 & 12 & Blue \\
Geometry & 25 & 7 & Target \\
Data Type & 21 & 9 & Aerial Imagery \\
Map Style & 21 & 9 & Artist Rendering \\
Map Type/Capability & 20 & 7 & Interactive Map \\
Administrative & 17 & 10 & Administrative Boundary \\
Basemap/Background & 17 & 7 & Basemap \\
Infrastructure & 17 & 7 & Bridge \\
Environmental Impact & 15 & 13 & Abandoned Drums \\
Civic & 13 & 8 & Airport \\
Area & 12 & 4 & Footprints \\
Lot/Land/Tax & 12 & 6 & Cadastral Data \\
Water & 12 & 9 & Constitution Lake \\
Greenspace & 11 & 6 & Trees \\
Prison & 11 & 6 & Adult Prison \\
Danger & 10 & 7 & Shooting Range \\
Government & 7 & 4 & Government \\
Trail & 6 & 5 & South River Trailhead \\
Demographics & 5 & 3 & Census \\
Park/Forest & 5 & 3 & Conservation Area \\
Clip Art & 4 & 4 & Clapboard \\
Commercial & 3 & 2 & Blackhall Studios \\
Facility & 3 & 2 & Cop City \\
Land Use & 2 & 2 & Housing \\
Addresses & 1 & 1 & Home Locations\\
 \hline
\end{tabular}
\end{centering}
\end{table}
}

\subsection{Mapping Strategies and Technology}

\subsubsection{Official Map}
\begin{figure}[h]
\centering
 \includegraphics[width=12cm]{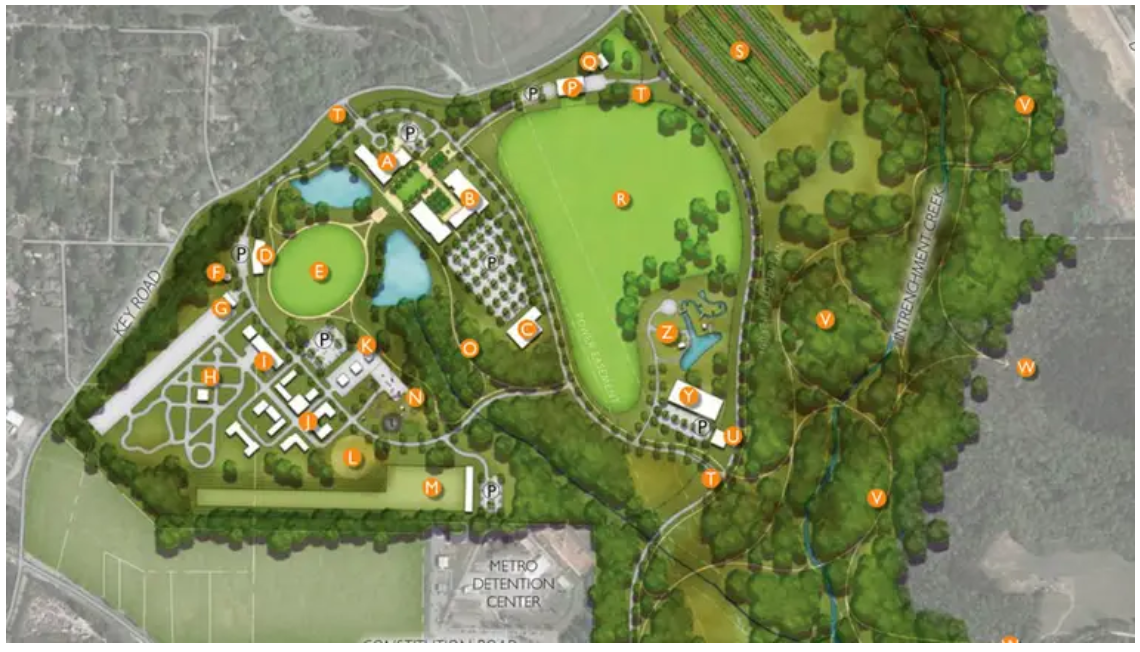}
 \caption{The Official Public Safety Training Center Plan. Source: \textit{Atlanta Police Foundation}, No date. This image corresponds to map 32 in Table \ref{Tab:DataSources}.}
 \label{APF_Map} 
\end{figure}

A map of the facility \textcolor{black}{which we refer to as the `official map'} is shared by the Atlanta Police Foundation. It has hand-drawn artistic elements that are rendered to appear 3D, including lush green trees and green areas, buildings, trails and water bodies (Figure \ref{APF_Map}). The map does not contain the traditional map elements described above, but has labels for Key Road, the Metro Detention Center and Intrenchment Creek. The trees are made to appear textured and almost `fluffy', to give the map a soft aesthetic. The local area is faded to de-emphasize the surroundings. These elements work to draw attention away from the local residents who are affected by the site, and instead, highlight the trees and greenspace in the facility. This map was duplicated frequently in news articles, reposted with annotations by individuals, and appeared often in our searches.

\textcolor{black}{The popularity of this `official map' released by the City of Atlanta, has two major implications. First, it implies lack of access to mapmaking tools or need for custom site maps for the general public, as they often rely on this map. Second, it implies a lack of understanding of the subtle ways this map design serves to support construction of the site.}

\subsubsection{Articulating the Plan}
\begin{figure}[!tbp]
 \begin{centering}
 \includegraphics[width=15cm]{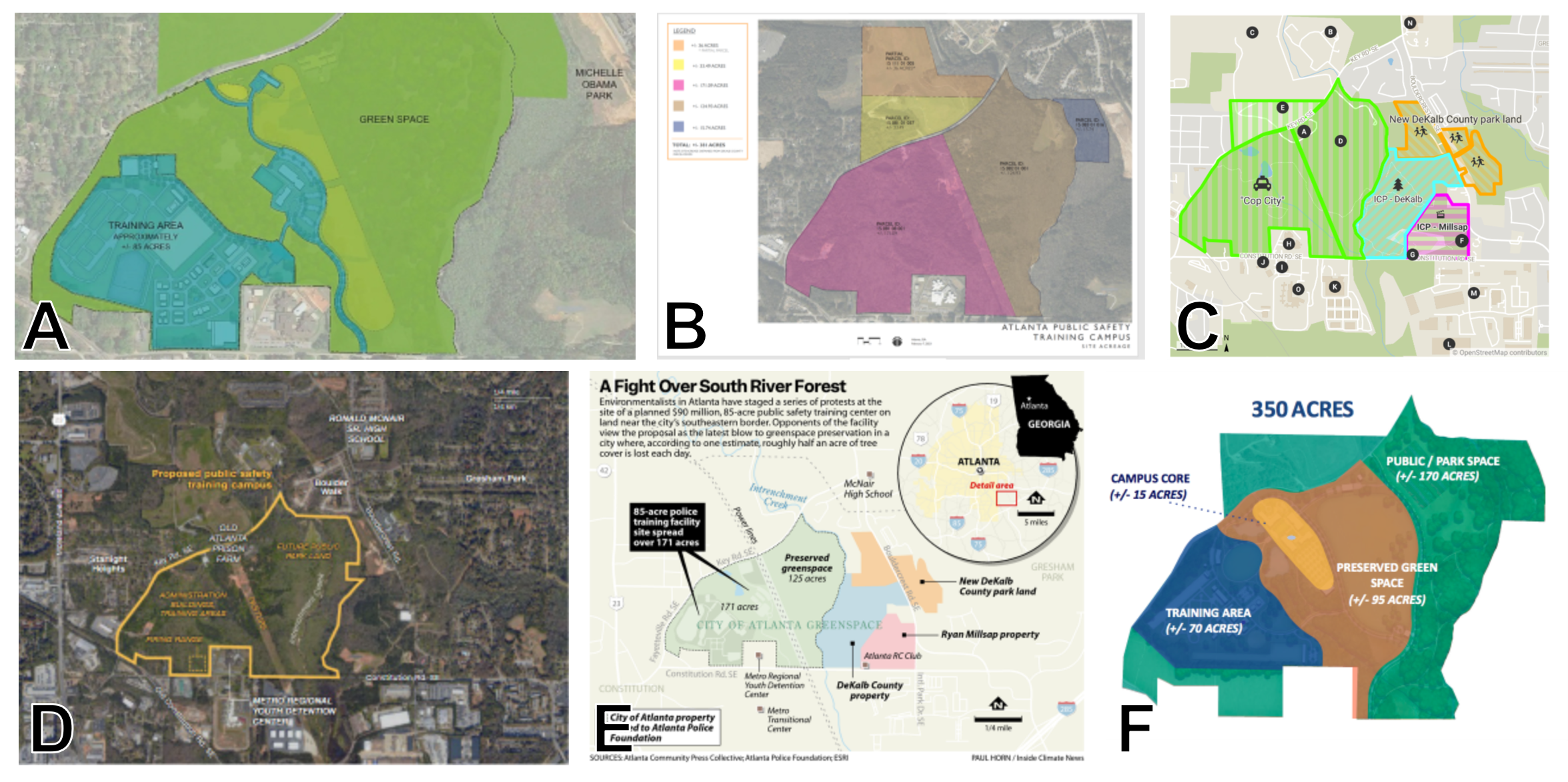}
 \caption{Map images using color-coded parcel selections to depict the site boundaries and land parcels involved in the project. These sections reference the training center space (A, C, D), the remaining public park space (A, B, D, F), property owned by Ryan Milsap (C, E). Some maps also offer the sizes of various parcels (B, E, F). Images A-F correspond to maps 18, 20, 21, 17, 23, and 5, respectively, in Table \ref{Tab:DataSources}.}
 \label{articulatingtheplan}
 \end{centering}
\end{figure}

The most common theme was the inclusion of polygons that showed the division of sections within the area (Figure \ref{articulatingtheplan}). Many maps depicted the outline of the boundary (or the polygons of the boundaries) of the area that the city owns, and which of the county's land parcels are included within the roughly 350 acre area. These maps conveyed that the training facility would be built on the west side of the parcel and the east side would be preserved as a green space. Maps included building and impervious surface footprints of site elements, such as the jogging track, to model what is proposed to be built inside the plot (Figures \ref{articulatingtheplan}A and \ref{articulatingtheplan}E). 

An article published by \textit{CNN} includes a map with labels of the firing range, youth detention center, old Atlanta prison farm, and high school (Figure \ref{articulatingtheplan}D), which allows readers to see the \textit{interaction} of features associated with violence, policing, and minors. The name `Millsap', appeared on two maps; Ryan Millsap, the Chairman \& CEO of Blackhall Studios, owns a portion of the land and has been embroiled in a longstanding contract issue with the site. We found no common standard for a map of this plan vis-a-vis land parcel ownership; conversely, it appears that different media outlets were creating their own versions of this map. 

Maps created by the City of Atlanta emphasize that some land will be preserved for public green space, while maps made to challenge the facility may include affected schools and residential buildings, and the loss of public green space.

\subsubsection{Use of Basemaps and Datawrapper}
\begin{figure}[!tbp]
 \begin{centering}
 \includegraphics[width=14cm]{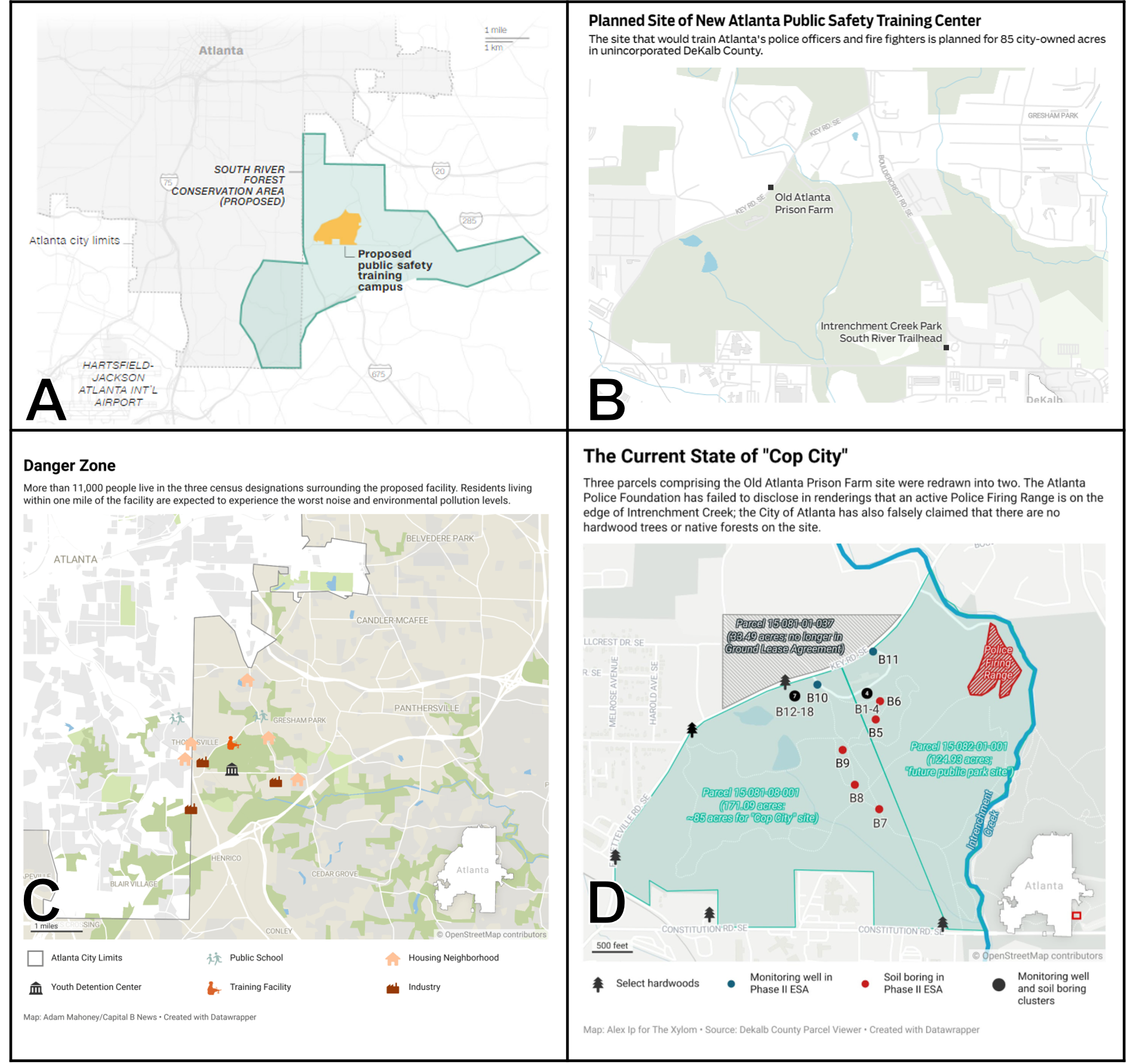}
 \caption{Maps created using Datawrapper, a publicly available software for data visualization. These maps present various placemarkers and environmental features (A, D), historic locations such as the Old Atlanta Prison Farm (B), and youth and family-related points of interest such as schools and housing neighborhoods (C). These maps also use parcel selections to depict the location in relation to the Atlanta city limits (A, C, D). Images A-D correspond to maps 16, 15, 13, and 28, respectively, in Table \ref{Tab:DataSources}.
 \label{Datawapper}}
 \end{centering}
\end{figure}

We estimate that up to eight maps were likely created using Datawrapper (four are shown in Figure \ref{Datawapper}), a free, online visualization software that is popular among leading news outlets (e.g., \textit{The Washington Post} and \textit{The New York Times}), and uses sources from OpenStreetMap. Datawrapper provides basemaps, which include predetermined elements, such as roads, rivers, labels, and colors, that users can automatically add elements or draw on. The mapmaker does not often control what is put on the map (although they can add elements to the map), whereas, traditionally, mapmakers trained in GIS and cartography tend to add features one-by-one and choose custom label fonts and styles. Conversely, maps used in the EIS documents and in official legal documents from the APF used professional GIS software that allows for statistical analysis (such as Esri ArcMap, which was likely used to create Figure \ref{articulatingtheplan}B).

Figures \ref{Datawapper}A and \ref{Datawapper}B from \textit{CNN} and the \textit{AJC}, respectively, appear in articles that describe the site and the controversy surrounding the project. Figure \ref{Datawapper}B is part of an article entitled ``‘Forest defenders’ use extreme tactics in fight over training center'' that describes the SCC opposition as violent and extremist.\footnote{The browser tab for the article is titled: ``Far-left activists use vandalism to protest Atlanta police training facility.''} Accordingly, the map has less context than similar maps that were generated using Datawrapper. 

Figures \ref{Datawapper}C and \ref{Datawapper}D were created by mapmakers who openly oppose the project. An map published in \textit{Capital B News}, a local non-profit news media that is centered on Black voices and issues\footnote{Capital B News: https://capitalbnews.org/about-us/}, calls the site a `danger zone' (Figure \ref{Datawapper}C) uses red shades, and includes icons of schools, housing, and industry. Figure \ref{Datawapper}D is part of a longer expos\'e accusing the city of making false claims \textcolor{black}{surrounding the project, e.g., that the buildings will not be built on forested land,} and pinpoints environmental monitoring and soil boring (i.e., sampling) sites, as shared in the EIS documents, as part of the narrative.

\textcolor{black}
{This use of DataWrapper is in contrast to many maps that were created by using existing maps and adding simple text or annotations which could be done using any image editing software. While some mapmakers who used Datawrapper openly oppose the project, it is more common among news publications that aim to appear neutral or oppose the project. This demonstrates that mapmaking software may be unknown and/or inaccessible to the wider community of citizens and activists engaged with this project.}

\subsubsection{Use of Color}
\begin{figure}[!tbp]
 \begin{centering}
 \includegraphics[width=15cm]{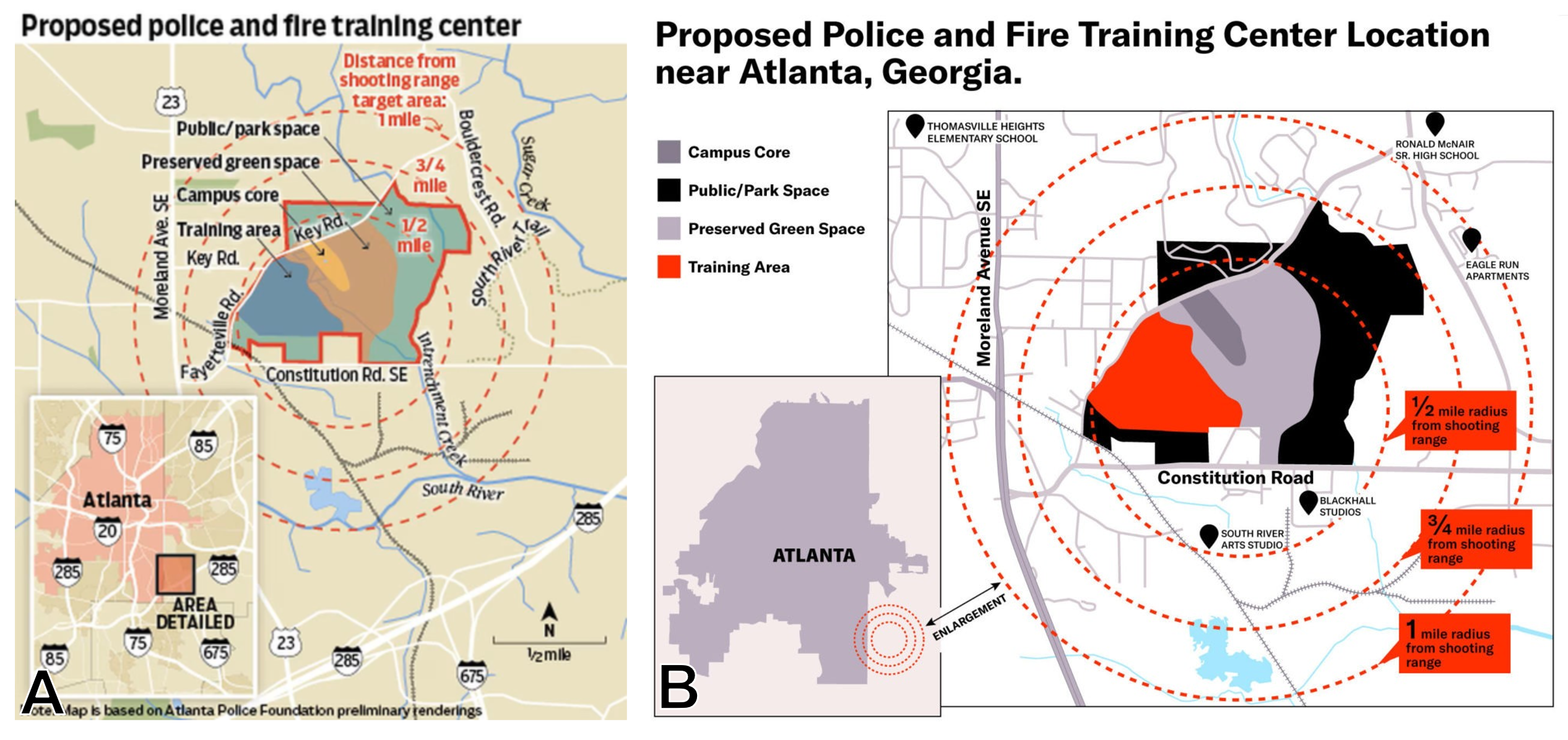}
 \caption{Maps from the (A) Atlanta Journal Constitution and (B) The Appeal each contain similar elements, style, and layouts. Images A and B correspond to maps 6 and 7, respectively, in Table \ref{Tab:DataSources}.
 \label{Provinance}}
 \end{centering}
\end{figure}

Maps can send different messages by using different color palettes \cite{anderson2021affective}. In our analysis, maps often used blue (25\% of maps) and green (25\%) which align with the colors of the natural features of the site. Other maps used red (37.5\%) which can evoke alarm or signify danger. A map published in 2021 in the \textit{Atlanta Journal Constitution} (\textit{AJC}) uses APF data and was likely made with Datawrapper (Figure \ref{Provinance}A). Four months later, a map using the same elements, inset map location, and distance target from the shooting range was published in \textit{The Appeal}, a venue dedicated to exposing systemic racism and supporting criminal justice reform\footnote{The Appeal: https://theappeal.org/about-us/} (Figure \ref{Provinance}B). Each map includes the shooting range target (in red), which communicates potential dangers to the area. The map in figure \ref{Provinance}B emphasizes danger and alarm by using red and black, and adds the elementary school, high school, and a local apartment complex to show readers that children and residents may be affected by the shooting range. The map in figure \ref{Provinance}A excludes these elements but includes labels of green space, waterways and roads.

\subsection{Concerns from the Stop Cop City Movement as Reflected in Maps}
In the following section we analyze the extent to which maps in our corpus reflect data related to key concerns of the Stop Cop City movement, including environmental justice, socioeconomic issues, civic infrastructure, and incarceration, violence and policing. Concerns of the movement that are not reflected in our corpus are discussed in section \ref{exclusion}.


\subsubsection{Elements Related to the Environment and Environmental Justice}
\begin{figure}[!tbp]
 \begin{centering}
 \includegraphics[width=15cm]{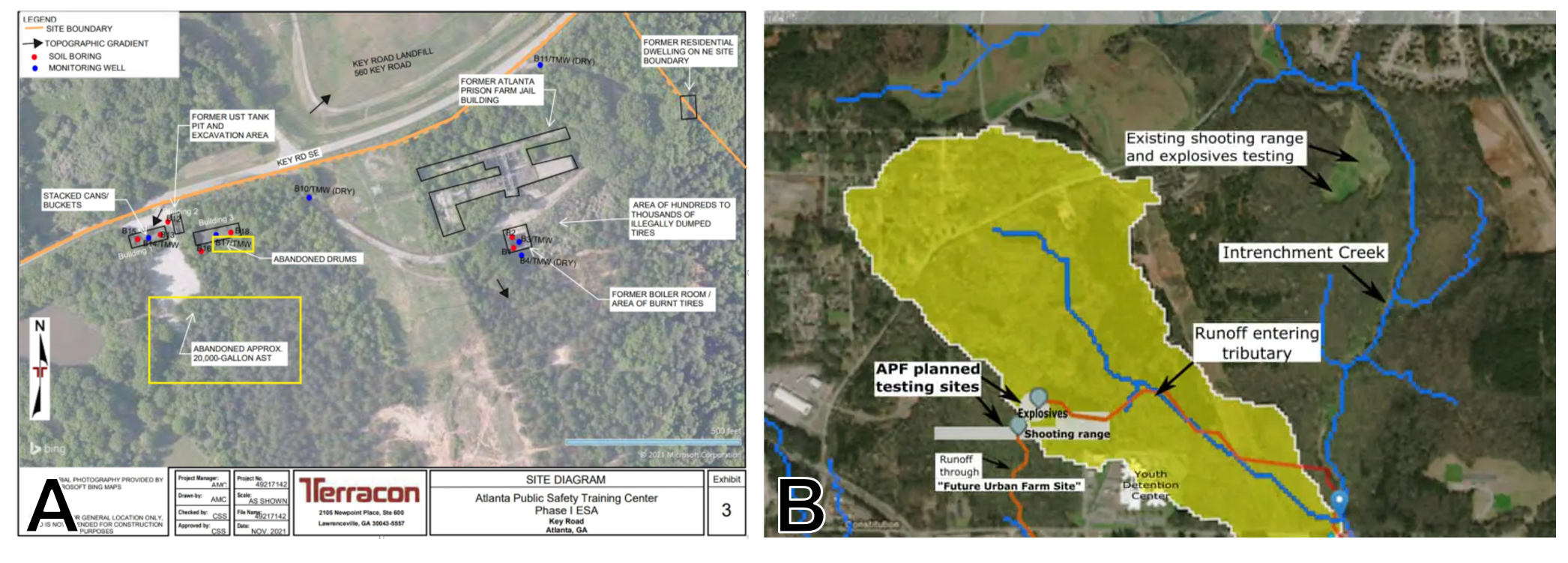}
 \caption{Maps depicting environmental information about the site, including (A) soil boring and monitoring wells, and (B) how future explosive and shooting range sites may produce runoff within the context of in the existing watershed. These maps correspond to maps 9 and 4, respectively, in Table \ref{Tab:DataSources}. 
 \label{Enviornmental}}
 \end{centering}
\end{figure}

Environmental features were sometimes illustrated on maps outside of the EIS documents. We retrieved maps from two articles, each of which was critical of the facility (a map from a third article, that was also critical, is displayed in Figure \ref{Datawapper}D). One article that opposed the facility published by the \textit{Atlanta Community Press Collective}, cites two maps created by Terracon for the EIS (one is shown in Figure \ref{Enviornmental}A) that show artifacts in the site. 
An article published in \textit{Mainlinezine} that describes environmental concerns about the facility contains a visualization that used United States Geological Survey (USGS) online modeling software StreamStats (Figure \ref{Enviornmental}B). StreamStats allows users to model hydrological features and visualize rasterized watersheds, runoff, and other water features atop aerial imagery. The mapmaker used the coordinates of the explosive sites and the shooting range and used the software to create a path of runoff (in red) from these origins. The map includes labels and a description saying that runoff from the planned explosives testing site and shooting range would enter the local tributary. This is the only example we found that uses a geocomputation model. This map also is a rare example of a mapmaker importing multiple (overlapping) data layers onto a map, outside of the maps found in the EIS documents. 

\subsubsection{Elements Related to Socioeconomic Issues}
\begin{figure}[!tbp]
 \begin{centering}
 \includegraphics[width=15cm]{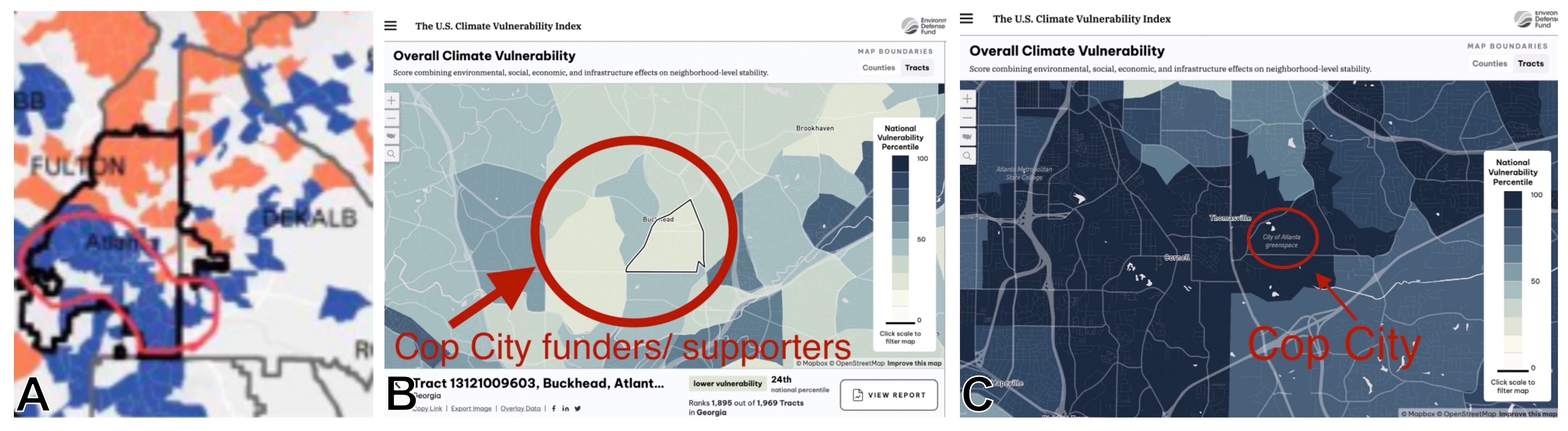}
 \caption{Maps depicting analysis of socioeconomic data of the neighborhoods relevant to the site. Map A includes the highest (orange) and lowest (blue) income areas in the Atlanta metro area using census tracts. Images B and C depict levels of climate risk with very high risk in dark blue and low risk in cream. Map B depicts the county where many supporters of the facility live, while map C depicts the county where the facility is slated to be built. Maps A-C correspond to maps 19, 30, and 29, respectively, in Table \ref{Tab:DataSources}.}
 \label{Demographics}
 \end{centering}
\end{figure}

We did not discover many maps on the topic of the facility that included socioeconomic data or mention of socioeconomic features. Such thematic maps are fundamental for GIS analysts to create (often with census data) to help viewers learn about who lives in certain areas. 
but, these data were not included in maps of the facility. Only mapmakers who appeared critical of the facility included socioeconomic data, and did so by annotating screen captures of maps that were built for other purposes (Figure \ref{Demographics}). 

One mapmaker took a screen capture of a map made by the Atlanta Regional Commission using census tract data of median household income, that was originally shared on the public media station \textit{WABE.org} (Figure \ref{Demographics}A). The mapmaker used this as a backdrop, drew a red line around low income tracts (in blue), and shared the map on Twitter, writing `\textit{`Both of the two bordering districts map onto far more neighborhoods in the bottom 20\% of median household income than the districts whose representation voted yes on Cop City.''} Using a similar strategy, Anti-Racist South added annotations to screen captures of an interactive tool that displays a climate vulnerability index at the census tract level (created by the Environmental Defense Fund and team) (Figure \ref{Demographics}B and \ref{Demographics}C). The mapmaker's goal was to show that climate vulnerability is lower in places where Cop City supporters live than where the facility is slated to be built. The tweet says,\textit{ ``Map 1: Atlanta neighborhoods adjacent to where they cut down 85+ acres of forest to build Cop City (91st, 94th \& 96th percentile of climate vulnerability). Map 2: Buckhead—where large percent of APF donors and politicians pushing Cop City live (24th percentile).''} 

\subsubsection{Elements Related to Local Residents and Civic Infrastructure}

In total, maps used elements related to civic features, including the airport, neighborhoods and schools, 13 times. In terms of related infrastructure, maps displayed features such as roads (n=11 appearances) most prominently, and some maps included and labeled utility lines (n=1) and bridges (n=2). Residences and housing were rare, and commercial or industrial firms were almost never included, with the exception of Blackhall Studios.

\begin{figure}[!tbp]
 \begin{centering}
 \includegraphics[width=6cm]{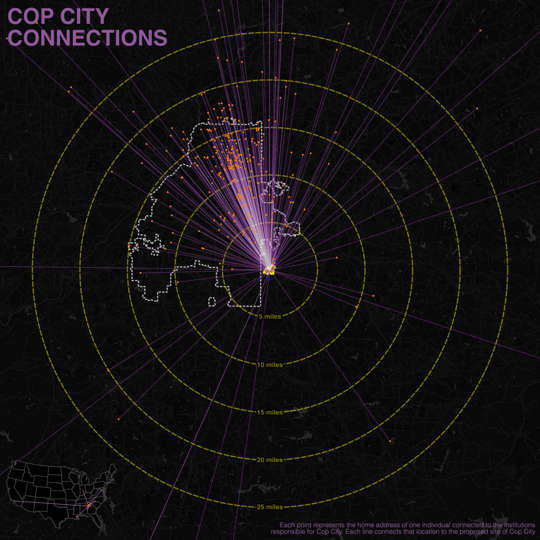}
 \caption{``Cop City Connections'' is a map connecting the home locations of those affiliated with the project to the site using lines from Taylor Shelton, \textit{Mapping Atlanta}. This image corresponds to map 27 in Table \ref{Tab:DataSources}.}
 \label{CopConnections}
 \end{centering}
\end{figure}

However, one map was designed to point out the homes of individuals. ``Cop City Connections'' is a map of the locations of individuals' homes who are ``\textit{connected to the institutions responsible for Cop City}'' (Figure \ref{CopConnections}). The mapmaker chose a black background and a target as stylist elements, and uses a network of nodes and edges to show connections of affiliates to the facility. The map uses cumulative negative space to direct the eye to a cluster of households that are northwest of the site. It provides little context, but instead focuses on the dispersion of the connections from the site. The map uses open source software QGIS.

\subsubsection{Elements Related to Incarceration, Policing, Violence, and Danger}
\begin{figure}[!tbp]
 \begin{centering}
 \includegraphics[width=15cm]{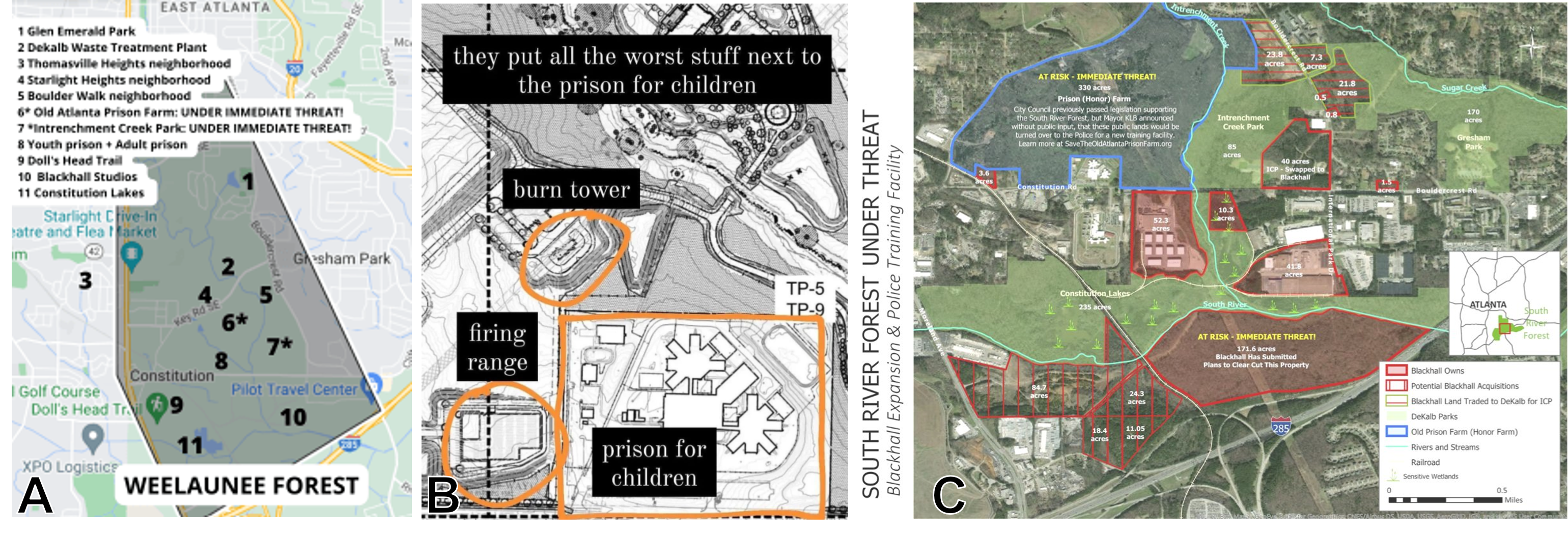}
 \caption{Several maps emphasized an immediate risk or danger to people or land. Maps A and C demonstrate the threat of deforestation in Intrenchment Creek Park and to the historic site The Old Atlanta Prison Farm. Map B emphasizes the threat to children, by highlighting the proximity of the burn tower and firing range of the proposed site to the existing child prison. Images A-C correspond to maps 14, 24, and 2, respectively, in Table \ref{Tab:DataSources}.}
 \label{Opposition}
 \end{centering}
\end{figure}

Narratives about abolition and over-policing in the SCC movement relate to physical features on, near, or in site plans, that signify incarceration and policing. Many maps point out landmarks related to incarceration (n=11). The Old Atlanta Prison Farm was the most prominently referenced landmark, and appeared on six maps, including maps by Terracon, the mass media, and the opposition. Maps also referenced the Youth Detention Center (n=3), which is also labeled as Prison for Children or Youth Prison (n=1 each) (Figures \ref{Opposition}A and \ref{Opposition}B). These two maps appear to have used other maps (such as Google Maps in Figure \ref{Opposition}A; origin unknown in Figure \ref{Opposition}B) for their arguments. The mapmakers appear to have annotated these images using other software that has free drawing features instead of making map labels, polygons, and annotations within the mapmaking software. 

Maps also contain elements \textcolor{black}{that} reflect a narrative of danger around the proposed facility (n=10 appearances). Maps in Figures \ref{Opposition}A and \ref{Opposition}C used the word `threat' to communicate that the planned site may be harmful. The shooting range is referenced five times: one map references an existing range (Figure \ref{Datawapper}D) and four others reference the range proposed in the facility plans (including those in Figure \ref{Provinance}). Most maps that reference the shooting range were part of narratives that opposed the facility.

\subsection{Exclusion} \label{exclusion}
Certain themes surrounding the SCC movement and the City of Atlanta's pro-facility narratives did not seem to appear in our map corpus. First, a focus of the Stop Cop City movement has been the preservation of indigenous land. The facility site was home to the Muscogee Creek people and is also called the Weelaunee Forest, an indigenous name \cite{pellow2023confronting}. We did not encounter reference to indigenous land on the maps, although one map by the opposition described labeled the site as Weelaunee Forest (Figure \ref{Opposition}A). Second, the City of Atlanta and the Atlanta Police Foundation used a narrative of crime reduction and safety to justify the building of this facility. However, we did not discover maps from these government entities that provided the public with geographic data about local crime rates or any positive impact on crime or benefits for the local community. Next, we found that the maps in our corpus do not communicate information about local businesses that may be affected by the facility and changes to the area. Finally, we also found no references to the number of people who live in the area (which can be defined using different areal metrics), and very limited representation of socioeconomic statistics about whom the facilitiy may affect.

\section{Discussion}
Maps are not to be understood as neutral communication devices, instead, they can have persuasive messages and can be used as part of an agenda or strategy. When exploring map products and mapping technologies related to the Atlanta Police Training Facility and the Stop Cop City Movement, we discovered a wide variety of mapping strategies, but the focus was typically on how the site was divided into areas (the training center, green space, etc.). We did not find evidence of spatial analysis or thematic mapping, and many of the maps' goals seemed to be to show the reader the plans for the land. 


As shown in the results, maps published by news media sources, including larger news conglomerates such as \textit{CNN} \textcolor{black}{(Figures \ref{articulatingtheplan}A, \ref{Datawapper}A)}, \textit{Atlanta Journal Constitution (AJC)} \textcolor{black}{(Figures \ref{Datawapper}B, \ref{Provinance}A)}, and \textit{NPR} (See S.I.), \textcolor{black}{did not appear to be designed to either support or oppose the facility. Such design choices may include displaying local crime statistics in an effort to argue for greater need for police to show favor for the facility or displaying civic elements or residencies that may be impacted by hearing gunshot sounds nearby or displaying the predicted increased flooding and exacerbated heat conditions to argue against it.} However, their companion articles sometimes negatively depicted protest actions. For instance, the \textit{AJC} (the source of reference 15 in Table \ref{Tab:DataSources}), is owned by Cox Enterprises which donated \$10 million to the APF and is considered a major backer of the facility. While their article is critical of the opposition, their map is relatively neutral. \textcolor{black}{This shows that even when more 'neutral' design decisions are used, maps can still be used to push for a specific viewpoint or agenda.}


We did not discover maps from individual users who were supportive of the facility, despite evidence that 39\% of polled Atlantans would vote in support of the facility \cite{Jilani_2023, Dandekar_2023}. Instead, many maps in the corpus were in support of the Stop Cop City movement. \textcolor{black}{This result echos prior research that shows that Far Right political groups in the U.S. who oppose the idea of climate change do not tend to make maps related to climate change, although climate change maps abound \cite{fish2023mapping}.}

We found that mapmakers aligned with the Stop Cop City movement often annotated aerial photos of pre-existing maps using digitally-drawn outlines, points, lines, etc. Data visualization research shows that the ability to draw on existing data visualizations helps creators convey a message \cite{ren2017chartaccent}. Here, the annotation process suggests that mapmakers may not have access to software or use software in the ways that professional mappers do. Instead, their annotations point the reader to either the location of an element related to key concerns, such as a shooting range or danger to the local community and neighboring child prison. They also point the reader to aspects of the population or area that they want to emphasize, such as differences in locale, poverty rate, and climate vulnerability between areas where supporters and funders of the facility live. This annotation method was especially popular with maps disseminated on social media (such as maps 2 and 24 in Table \ref{Tab:DataSources} and those in Figure \ref{Demographics}). Six annotated maps were not analyzed as part of the study but are included in the data corpus as examples. 

Digitally-drawn annotations to graphics can engage viewers and potentially increase their empathy (as seen with handwriting in \citet{tassiello2018handwriting}). This practice may also signify graffiti that challenges or defaces a graphic, or these homemade-style maps could be seen as less trustworthy as those created with tools designed for professional mapmaking. The maps made by the opposition (particularly those in Figure \ref{Demographics} and Figure \ref{Opposition}) would have likely looked different if the mapmakers used professional mapping software.

There was virtually no computation or fusion of datasets in this corpus, meaning that GIS operations were not leveraged to help individuals articulate their support or criticism of the facility. One exception is the maps that used the USGS StreamStats system (Table \ref{Tab:DataSources}, Maps 3 and 4), which exemplify how fusing GIS data layers and computing hydrological output using a back-end model can help explain how runoff affects the local environment and where the facility fits within a watershed. This is particularly valuable because it provides more context on how the new buildings will \textit{interact} with the ecosystem, instead of simply labeling map features. It also tells a story of cause-and-effect. Those who are aware of modeling resources and can leverage geocomputation and GIS operations may be able to tell a more detailed and scientifically-sound story to support their position. 

\textcolor{black}{Maps created in opposition of the facility and support of the SCC movement illustrate several key takeaways. First, the free-drawing annotations and reuse of other existing maps suggests that this group of mapmakers may be less aware of existing map-making software, lack training in these software systems, and/or as suggested by \citet{sieber2006public}, may have limited time to use these mapmaking tools effectively. 
This points to the need for a free, accessible, easy to use mapping software with low-barrier to entry for less conventional map-makers. Secondly, it shows that despite lack of access to these conventional tools, mapmaking is still made accessible through drawing and photo editing software, but limits these mapmakers to editing and annotating previously created maps, in this case, most often the `official map' created by the city of Atlanta. Subtle soft and `fluffy' design decisions of the official map emphasize claims that the facility is a positive addition to the Atlanta region, and minimize claims that oppose the facility. This shows the power that mapmakers can have in influencing public perception, even when activists or community members engage in direct counter-mapping to oppose that power. Ultimately, user-friendly mapmaking software might better support activists and the general public in visualizing, learning about, and communicating geospatial information.}

\begin{figure}
 \centering
 \includegraphics[width=15cm]{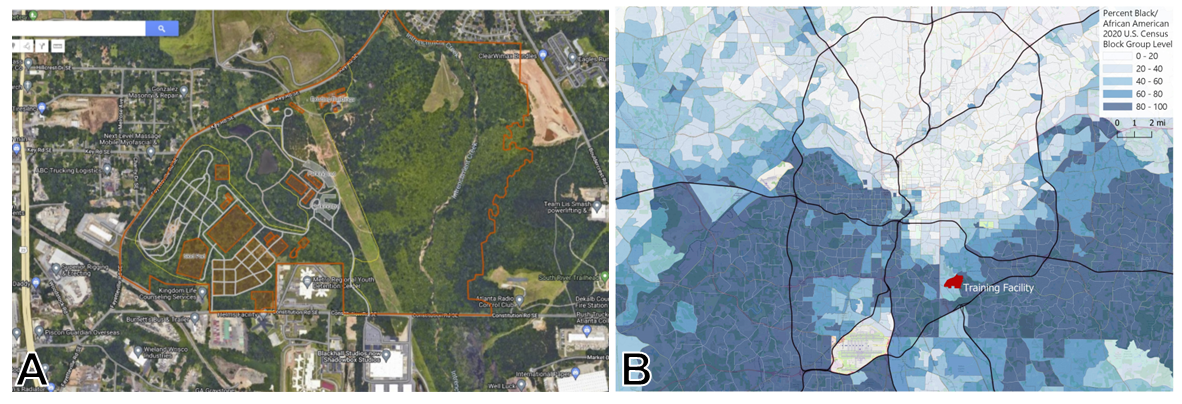}
 \caption{The authors created two maps to share information about the facility. Map A is a screen capture of an interactive map hosted by Google Maps that displays a keyhole markup language (KML) file of the planned facility infrastructure and boundaries. The KML data was digitized by the authors from official maps of the site plans (Figure \ref{APF_Map}). Map B is a tract-level choropleth map of Atlanta neighborhoods by percent Black or African American made in QGIS.}
 \label{fig:discussionpics}
\end{figure}

To demonstrate the capabilities of GIS technology as narratives, we designed two maps that display data that reflects key concerns (Figure \ref{fig:discussionpics}). Figure \ref{fig:discussionpics}A (link redacted for anonymity) is a screencap of a public, interactive Google Map where users can pan and zoom to learn about the surrounding commercial firms. Figure \ref{fig:discussionpics}B displays census tracts in the Atlanta metro area by percent Black population, with the facility highlighted in red. While Black Atlantans have been central to the Stop Cop City movement since it began, and increased violence towards Black residents continues to be a key concern, no maps in our corpus displayed data about race.

\section{Limitations and Future Work}
Our study has multiple limitations. First, this exploration does not measure the impact of the maps in our analysis. We did not track how often maps were shared or liked on social media, or the presence of the same map across multiple articles or posts. Further, we did not survey map viewers about their exposure to different maps. While our analysis offers a better understanding of mapping and counter-mapping strategies used by different stakeholders, it does not capture how frequently maps created by different stakeholders are seen by the public. Future work in this area could explore user perceptions or exposure to maps created by different stakeholders. Next, choosing keywords and higher-level themes for the maps (e.g, ``neighborhood'', ``highway'') was subjective and prone to human error. In the future, using a predefined set of keywords may eliminate semantic duplication, typographic errors, and help the coder avoid overly-coarse or fine-grained keywords. \textcolor{black}{We also hope to explore automated technologies to detect elements such as color, north arrows, scale bars and label text, and to capture differences in map style choices (e.g., the use of religious symbols to mark cemeteries), which can reflect value systems and local norms \cite{kent2009topographic}.} 

In many cases, maps appeared to use basemaps, which made it difficult to analyze which features and symbology were important to the mapmaker. As such, we were unable to confidently assess the mapmaker's choices. For instance, some basemaps showed Intrenchment Creek, but others did not; this may or may not have aligned with the mapmakers intentions as it may have been a built-in feature of the basemap. Thus, tallying and discussing the elements of maps that use a basemap does not tell us about the user's intention, per se, only the end product. 

In addition, this study covers a limited sample of maps (although we found many maps repeated in our multiple searches). The authors know of multiple maps and cartographic resources that were shared in February 2023 on Twitter, but due to the many SCC-related posts in 2023 and limited number of tweets returned for each of our keywords, these maps were not retrieved as part of our study. An ArcGIS Storymap describing the South River Forest was also not returned in Google searches and was also not included in our study for this reason. Accordingly, this case illustrates the need for academic researchers to have access to APIs for social media platforms. In the future, we plan to search for more UGC and examples from the EIS and other legal documents which are considered public information and are housed at the U.S. Library of Congress. 

We also did not interview mapmakers, which could yield better insights about their mapmaking process. In addition to talking to mapmakers, speaking with map viewers could help gauge the impact of these maps on public opinion. In the future, we could also speak with people involved in the facility's construction and in the opposition to learn about what types of maps and visualizations have been particularly salient, noteworthy, valuable, and memorable for them and their cause. We could further explore the topic of maps in the SCC movement by leveraging interview and survey methods to learn about different stakeholders' approaches to map-making and map viewing.

\section{Conclusion}
 In this manuscript, we searched Google Images, Google, Twitter, Instagram, and reddit to identify cartographic media on the topic of a proposed police training facility in Atlanta. \textcolor{black}{These maps may influence community members' opinions about the facility, and illustrate how different stakeholders communicate messages about the facility. We then conducted a content analysis of the collected images.} This exercise illustrated that the the government used lush renderings and artistic work and the media used a generally balanced mapmaking strategy, despite how the maps' companion articles were positioned. Counter-mappers in the Stop Cop City movement often used ad hoc screen capping and annotations of pre-existing map resources, although some activists and movement sympathizers used mapping technologies such as DataWrapper and QGIS. This exploration helped portray the variety of ways the public could engage with contentious site plans through maps, and how map style, elements, and technologies can influence how the public perceives a contentious site plan for a police training facility.

\section{Supplementary Material}
Supplementary material is available at \href{https://doi.org/10.7910/DVN/PCQ294}{https://doi.org/10.7910/DVN/PCQ294}.
\bibliographystyle{ACM-Reference-Format}
\bibliography{sample-base}









\end{document}